\documentstyle[prl,aps,multicol,epsf]{revtex}
\begin{document}
\draft

\title{3D~$XY$ and lowest Landau level fluctuations in 
deoxygenated YBa$_2$Cu$_3$O$_{7-\delta}$ thin films}
\author{Katerina Moloni, Mark Friesen, Shi Li, Victor Souw, P. 
Metcalf, Lifang Hou, and M.  McElfresh}
\address{Physics Department, Purdue University, West Lafayette,
IN 47907-1396}
\date{\today }
\maketitle
\begin{abstract}
Conductivity measurements reflect vortex solid melting in 
YBa$_2$Cu$_3$O$_{7-\delta}$ films.   Field-independent
glass exponents $\nu_g\simeq 1.9$ and $z_g\simeq 4.0$ 
describe the transition $T_g(H)$ for $0<H\leq 26$~T.  
At low fields, 3D~$XY$ exponents $\nu_{XY}\simeq 0.63$ and 
$z_{XY}\simeq 
1.25$ are also observed, with $z_{XY}$ smaller than  
expected.  These compete with glass scaling according to multicritical 
theory.  A predicted power-law form of $T_g(H)$ is observed for 
$0.5T_c< T_g<T_c$.  
For $T_g<0.5T_c$,  3D~$XY$ 
scaling fails, but a lowest Landau level analysis becomes 
possible, obtaining $T_{c2}(H)$ with positive curvature.  
\end{abstract}
\pacs{74.25.Bt,74.25.Dw,74.72.-h}
\begin{multicols}{2}

The nature of fluctuations near the superconducting to normal 
state transition in high-temperature 
superconductors (HTSCs) is still a matter of 
controversy.  Several distinct fluctuation types and regions have 
been 
proposed, {\it e.g.}, 3D~$XY$ fluctuations at low fields, lowest 
Landau level (LLL) fluctuations at high fields, and glass-like 
fluctuations 
(for disordered HTSCs) near the finite-field transition $T_g(H)$.  
However, experimental analyses based upon the different 
scaling theories lead to conflicting results.  This situation is 
most evident for competing 3D~$XY$ and LLL fluctuations, {\it both} of 
which are supported experimentally, in the same region of the phase 
diagram, in spite of being incompatible \cite{lawrie}.

The 3D~$XY$ transition is driven by phase fluctuations of a complex 
order parameter (OP), which fall into the universality class of the 
$\lambda$ transition in $^4$He.  The zero-field, ``intermediate" 
(nonelectrodynamic) phase fluctuations of the HTSCs are thought to 
be of this type \cite{FFH}.  At $T=T_c$ (and $H=0$), these fluctuations 
diverge in size, driving the resistive phase transition.  Recent 
experimental evidence supporting this picture is found 
in specific heat \cite{inderhees,salamon}, magnetization 
\cite{salamon,hubbard}, penetration depth \cite{kamal}, 
and current-voltage ($J$-$E$) measurements \cite{salamon,yeh}.  
The finite-field transition $T_g(H)$, which is similarly 
driven by phase 
fluctuations of the OP, joins smoothly to $T_c\equiv T_g(H=0)$.  
However, the glass and 3D~$XY$ fluctuations exhibit distinct scaling 
functions and exponents \cite{FFH}.

Fluctuations of the OP {\it amplitude} occur near the upper critical 
(mean-field) temperature $T_{c2}(H)$.  These fluctuations drive the 
Cooper pair 
density to zero, but do not correspond to a true transition;
superconducting order vanishes at the slightly lower temperature 
$T_g(H)$.
In the low-field region, the distinction between OP amplitude and 
phase fluctuations results in the dominance of 3D~$XY$ critical 
behavior.  At high fields, this distinction is not present, 
yielding a different type of behavior, most conveniently described in 
terms of the Ginzburg-Landau LLL approximation, with its 
corresponding scaling theory \cite{bray,ullah}.  
Experimental evidence in support of this behavior is found in
specific heat \cite{welp,zhou,jeandupeux}, 
magnetization \cite{welp,jeandupeux},
and $J$-$E$ characteristics \cite{welp,kim}.  A crossover is 
expected between the low-field (3D~$XY$) and high-field (LLL) 
behaviors, and its 
clarification is fundamental in the investigation of HTSC fluctuations 
\cite{zlatko}.

It may appear that glass fluctuations near
$T_g(H)$ only complicate the story, since they compete with both the 
low- and high-field 
fluctuations.  However, in this 
letter we suggest, to the contrary, that glass fluctuations {\it help} 
to identify low- and high-field behaviors, through the use of 
multicritical scaling theory.  
Such theories   
are applicable when fluctuations of different types compete 
for dominance.  For example, in the low-field limit, $T_g(H)$ {\it 
joins} $T_c\equiv T_g(H=0)$, forcing the distinct fluctuation types to 
coexist near the multicritical point $T_c$ \cite{FFH}.  In this paper, 
multicritical 
predictions augment the 3DXY and LLL 
theories, thus clarifying their applicability to HTSCs.

To address these issues, it is desirable to work in both the low- and 
high-field regions of the phase diagram.  
In YBa$_2$Cu$_3$O$_{7-\delta}$ samples, $H_{c2}(0)$ is very large
($\gtrsim 100$~T)
for ``optimal" ($\delta \simeq 0.05$)
stoichiometry.  We therefore focus on deoxygenated 
films, for which magnetic field scales 
are relatively small, allowing $H_{c2}(T)$ to be accessed over a wide 
temperature range.  Three ``optimal" $c$-axis 
YBa$_2$Cu$_3$O$_{7-\delta}$ films, 
approximately 4000 \AA  ~thick, 
were prepared, and then deoxygenated, as described elsewhere 
\cite{hou}.  Films with 
resulting stoichiometries 
of $\delta \simeq$ 0.24, 0.57, and 0.59 ($\pm 0.05$)
were produced, corresponding to $T_c$s of 
77, 61, and 56~K, respectively.  
The films were patterned into 100 $\mu$m $\times$ 2000 
$\mu$m bridges using laser ablation.  Isothermal $J$-$E$ 
and conductivity ($\sigma$)
curves were obtained using a conventional four-point geometry.  
As typical for underdoped samples, the normal state contribution 
could not be eliminated from $\sigma$, as it is not yet
well characterized.  However, this background contribution should 
not affect the present results greatly, in the temperature range of 
interest, due to the insulating nature of the normal state.
Magnetic fields were applied perpendicular to the film 
surface (parallel to the $c$-axis) in the range $0< H\leq 26$ T.

Transition temperatures $T_g$, corresponding to the continuous 
vortex solid 
melting transition, were deduced at each applied field using 
the scaling ansatz of Fisher, Fisher, and Huse (FFH) \cite{FFH}.  (Note that 
this ansatz is isomorphic with any two-exponent scaling theory of the melting 
transition.)
The appropriately scaled conductivity is 
given by $(J/E)|T-T_g|^{\nu_g(z_g-1)}$, 
while the scaled current is 
$(J/T)|T-T_g|^{-2\nu_g}$.  Here, $\nu_g$ and $z_g$ are the 
static and dynamic glass scaling exponents, respectively.  The 
$H_g(T)$ phase boundaries 
determined in this way are shown in Fig.~1.  The 
obtained scaling 
exponents are independent of $\delta$ \cite{hou} and $H$, with 
values of $\nu_g =$ 1.8-1.95 
and $z_g =$ 4.0-4.1.  This is consistent with the notion of a single, 
3D, glass universality class.
 
As a first application of multicritical scaling, we note that the form of 
$H_g(T)$ [the inverse of $T_g(H)$] is specified at low fields by the 3D~$XY$ 
theory \cite{FFH,salamon}:  $H_g(T)=H^*(1-T/T_c)^{2\nu_{XY}}$, where 
$\nu_{XY}$ ($\neq \nu_g$) is the 3D~$XY$ static exponent.  The 
phase boundaries 
satisfy this relation over a wide temperature range, 
$0.5T_c\leq T\leq T_c$, as shown in Fig.~1 (inset), identifying  
the crossover temperature $T_b\simeq 0.5T_c$ as 
the limit of 3D~$XY$ scaling.  The 
slope of these curves gives a (sample-averaged) exponent 
$\nu_{XY}=0.63\pm 0.04$, which compares favorably with the 
expected value of 0.669 \cite{zinn}.  This is then used to determine 
the 
sample-dependent field scale $H^*$.  For the films used here, $H^*$ 
is in the range 7-19~T,
as compared to ``optimal" samples, for which 
$H^*$ is on the order of 50 T.  The crossover field 
$H_b\equiv H_g(T=T_b)$ was 
studied previously, and was suggested to separate 3D from 
2D behavior along $H_g(T)$ \cite{deak,schilling}.  However, the 
field-independence of the glass transition does not corroborate this 
conclusion.  Below, we 
demonstrate instead that for fields $H<H_b$, the 3D~$XY$ 
description is in good agreement with the data, while for $H>H_b$, a 
self-consistent LLL description becomes possible.

It is possible to determine $T_c$ 
at zero field, using the FFH ansatz \cite{FFH}.  However, this 
procedure is known to become uncontrolled at very low 
fields, obtaining surprising results, such as nonuniversal exponents 
\cite{roberts}.  Therefore, we develop here a more reliable 
``crossing point" scaling technique, by extending the 3D~$XY$ analysis 
to finite fields \cite{friesen}.  The subsequent field 
scaling hypothesis involves
the 3D~$XY$ scaling variable \cite{FFH,salamon}:  
$x=(H^*/H)^{1/2\nu_{XY}}(T-T_c)/T_c$.  The 
scaling of 
the ohmic conductivity $\sigma_\Omega$ can then be written as
$\sigma_\Omega (H/H^*)^{(z_{XY}-1)/2} =\tilde{s} (x)$, for which 
the asymptotic behavior is known 
\cite{salamon,friesen}:  
$\tilde{s} (x)\sim (x+1)^{-\nu_g(z_g-1)}$ 
as $x\rightarrow -1$, corresponding to $T\rightarrow T_g(H)$; and 
$\tilde{s} (x)\sim x^{-\nu_{XY}(z_{XY}-1)}$ as 
$x\rightarrow +\infty$, corresponding to $H\rightarrow 0$.

The crossing point method proceeds from the definition of $x$:  if 
$H>0$ and $T=T_c$, then $x=0$, and must therefore be 
independent of $H$.  It follows that in this limit,
 $\sigma_\Omega (H/H^*)^{(z_{XY}-1)/2}$ should also be 
independent of 
$H$.  Data sets $\sigma_\Omega (T)$, obtained at constant fields 
$H$, must then all cross at $T=T_c$ when plotted as 
in Fig.~2, provided that 
$z_{XY}$ has been chosen correctly.  As observed in the insets, the 
crossing point method 
places strong constraints on the exponent and transition 
temperature, 
which we identify as $z_{XY}=1.25\pm 0.05$ and 
$T_c=60.8\pm 0.4$, 
for the film shown.  These results are corroborated by low field data 
($H\leq 0.1$ T) \cite{friesen}.  Since 3D~$XY$ fluctuations are most 
prevalent near $T=T_c$,
it is helpful to think of this method as  
optimizing scaling near this temperature.  We emphasize that 
$z_{XY}$ obtained in this 
way disagrees with the expected \cite{FFH} diffusive dynamics ($z_{XY}=2$), 
and also with other 
experimental observations \cite{roberts,anlage}.  It is our opinion that the 
crossing point method can achieve better results than those analyses, due to
the incorporation of finite field data in the scaling procedure.

The estimates for $z_{XY}$, $\nu_{XY}$, and $T_c$, determined 
above, can be checked through a full scaling analysis of 
$\sigma_\Omega$, as shown in Fig.~3.  For that film, the initial 
estimates
could not be improved upon.  
The fitting is excellent at all low fields, becoming optimal at $T=T_c$.  
For comparison, the best fit, using the 
expected exponents $z_{XY}=2$ and $\nu_{XY}=0.669$, is shown in 
the inset.  As found in previous analyses \cite{salamon}, scaling using 
$z_{XY}=2$ suffers at the lowest fields, and perhaps more 
importantly, 
near $T_c$.  
The scaling results found here are therefore an improvement over 
previous analyses.  

The crossover between low- and high-field behavior, observed above 
using multicritical scaling, can be formulated more conveniently as 
follows.  Since all sample dependence of the 3D~$XY$ 
scaling variable $x$ is absorbed into the characteristic field $H^*$, the 
divergence of $\sigma_\Omega$ in Fig.~3, must 
occur at a universal value of $x=-1$ \cite{friesen}.  Multicritical 
self-consistency  therefore requires the scaling variable 
$x_g$ [{\it i.e.,} $x$ evaluated along the phase boundary $T_g(H)$] 
to remain field-independent in the 3D~$XY$ scaling 
region, as
shown in the lower half of Fig.~4.  Errors in the determination of 
$x_g$ are magnified at lowest fields, where the difference 
$T_c-T_g(H)$ is small.  Deviation from 3D~$XY$ multicritical 
self-consistency 
becomes apparent for $H>H_b$.

Several fluctuation models are candidates for describing the upturn 
of $H_g(T)$ when $H_g>H_b$ in Fig.~1 
(inset).  Here, we consider the 3D~LLL model, using
multicritical theory to place restrictions on the allowable 
scaling.  The scaling technique is 
constructed in 
analogy with the preceding 
3D~$XY$ analysis.  In the LLL theory, a natural scaling parameter 
emerges \cite{bray,ullah}:  $y=(H^*T_c/HT)^{2/3}[T-
T_{c2}(H)]/T_c$, where $T_c$ 
and $H^*$ have been used here to make $y$ dimensionless.  
In this 
analysis it is $T_{c2}(H)$ which must 
be determined by 
scaling.  Although $T_{c2}(H)$ is often assumed to be linear 
\cite{bray,ullah}, this restriction becomes too severe for the underdoped 
samples used here.  Instead, a LLL crossing point method is now 
constructed, which allows the {\it form} of $T_{c2}(H)$ to be 
ascertained.  As 
described below, this analysis obtains $T_{c2}(H)$ curves with 
positive curvature---an interesting feature which has previously 
been 
associated only with magnetically doped HTSCs \cite{kresin}.

The LLL crossing point method is described as follows.  We make use 
of the 3D~LLL scaling ansatz \cite{ullah}, which we write as 
$\sigma_\Omega (HT_c^2/H^*T^2)^{1/3}=F_{\rm 3D} (y)$.  
In analogy with 
the 3D~$XY$ case, $\sigma_\Omega (HT_c^2/H^*T^2)^{1/3}$ must 
be 
independent of $H$ when $T=T_{c2}(H)$.  To begin, a value of 
$T_{c2}$ is first {\it assumed} for a particular reference field $H_0$.  
The temperatures $T_{c2}(H)$ consistent with this choice are then 
obtained for other fields ($H\neq H_0$).  This is accomplished by 
plotting 
$\sigma_\Omega (HT_c^2/H^*T^2)^{1/3}$ {\it vs.} $T-T_{c2}(H)$  
for (fixed $H$) $\sigma_\Omega (T)$ data sets, 
then adjusting $T_{c2}(H)$ for each field, until a crossing occurs at 
$T-T_{c2}(H)=0$, similar to Fig.~2.  For 
fields $H\neq H_0$, $T_{c2}(H)$ is thus a 
function of the original choice of $T_{c2}(H_0)$, reducing the 
following fit to a single parameter.  All $T_{c2}(H)$ curves found in 
this way exhibit positive curvature.  It is once again helpful to view 
the crossing point as a method for optimizing scaling in the most 
essential 
temperature region; in the LLL case this is near 
$T_{c2}(H)$.

We are now left with one fitting 
parameter, $T_{c2}(H_0)$, 
which cannot be estimated from the crossing point 
method alone.  The full LLL scaling procedure is now used to 
determine the single fitting parameter, while simultaneously 
requiring multicritical 
self-consistency.  In analogy with the 3D~$XY$ case, this means that 
the scaling variable $y_g$ [{\it i.e.,} $y$ evaluated at $T=T_g(H)$]
must remain field-independent in the LLL scaling region.  We find 
that by 
using $T_{c2}(H)$ obtained from the crossing point method, 
multicritical  self-consistency {\it cannot be met at low fields}.  Since the 
LLL theory has greatest justification at high fields, we attempt, 
instead, 
to obtain self-consistency in the high-field region.  
The outcome of the final scaling procedure is shown in Fig.~5.
The (approximate) field-independence of $y_g$ is evident 
for the entire high-field range $H>H_b$, as 
shown in the top half of Fig.~4.  $T_{c2}(H)$ is obtained with only small 
uncertainty, 
as shown in Fig.~5 (inset).

We comment finally on the difference between the present results 
and those of Refs.~\cite{welp} and~\cite{kim}.  In our work, LLL 
scaling is found to be multicritically self-consistent only at high fields 
($H>H_b$), while Refs.~\cite{welp,kim}, which do not check for 
self-consistency, find that LLL scaling 
is applicable at low fields ($H<H_b$).  (Note that $H_b$ is very large 
in the ``optimal" samples used by those authors.)  We speculate that 
scaling could be accomplished in Refs.~\cite{welp,kim} only by 
allowing diminished scaling quality in precisely the region 
where the quality should be highest 
[{\it i.e., } near $T_{c2}(H)$].  In the present 
work, this 
situation is avoided by optimizing scaling near $T_{c2}(H)$ from the 
outset.  

After the completion of this work, we learned of similar 
3D~$XY$ results, obtaining $z_{XY}\lesssim 2$ from conductivity 
scaling \cite{urbana}.

We thank S.~Girvin, A.~MacDonald, S.~Pierson, Z.~Te\v{s}anovi\'{c}, 
and especially P. Muzikar for many 
helpful discussions.
This work was supported through 
the Midwest Superconductivity Consortium (MISCON) DOE grant \# 
DE-FG02-90ER45427, the Materials Research Science 
and Engineering Center (MRSEC) Program of the NSF under Award \# 
DMR-9400415, and the National High Magnetic Field Laboratory 
(Tallahassee).

\begin{figure}
\narrowtext
\epsfxsize=2.9truein
\vbox{\hskip 0.15truein
\epsffile{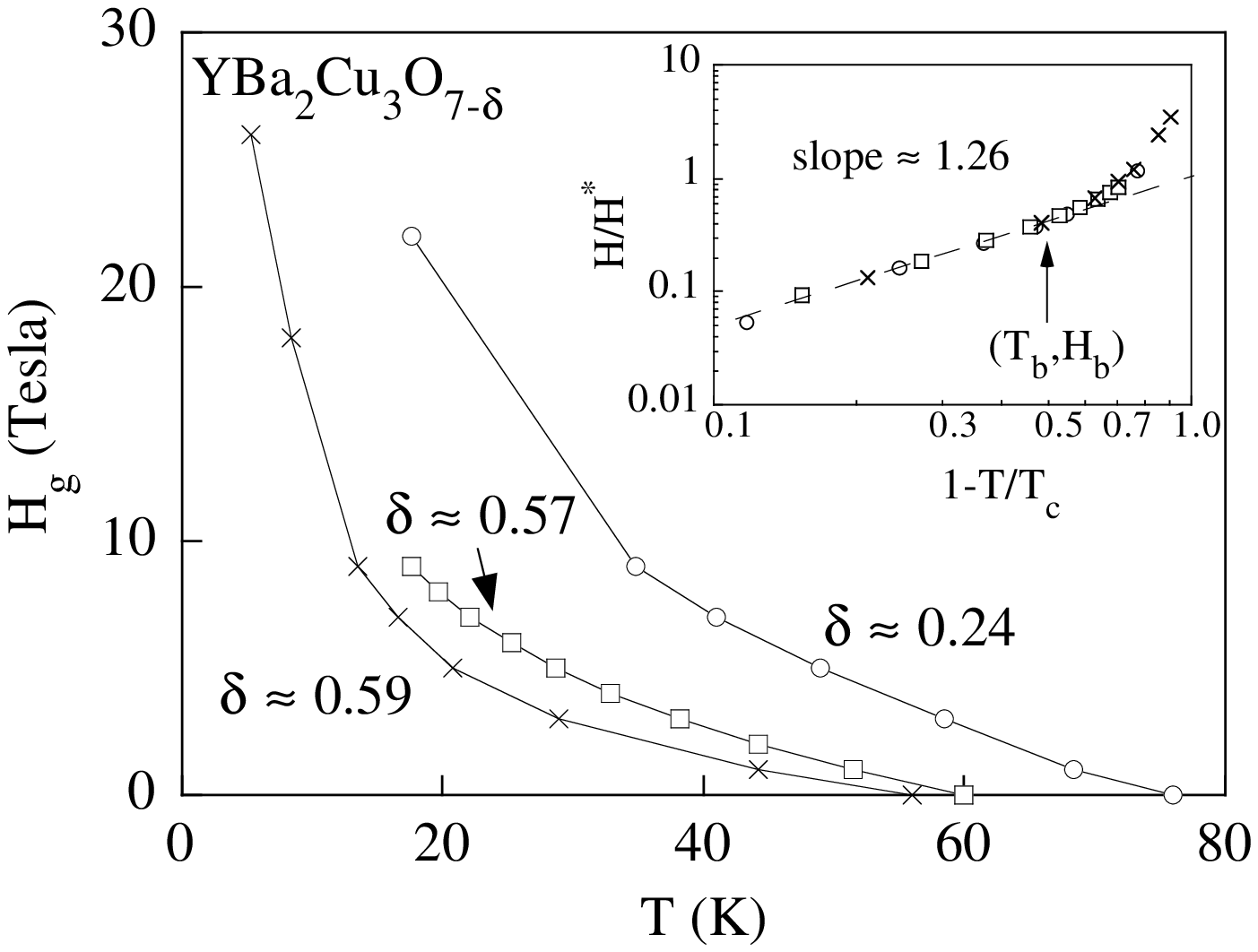}}
\end{figure}
\vspace{-.4in}
\noindent FIG.~1. Superconducting phase 
boundaries for three deoxygenated YBa$_2$Cu$_3$O$_{7-\delta}$
thin films.  Lines are a guide to the eye.
Inset shows the same phase boundaries plotted logarithmically.  
Power law behavior is evident for $H<H_b$.

\begin{figure}
\narrowtext
\epsfxsize=2.9truein
\vbox{\hskip 0.15truein
\epsffile{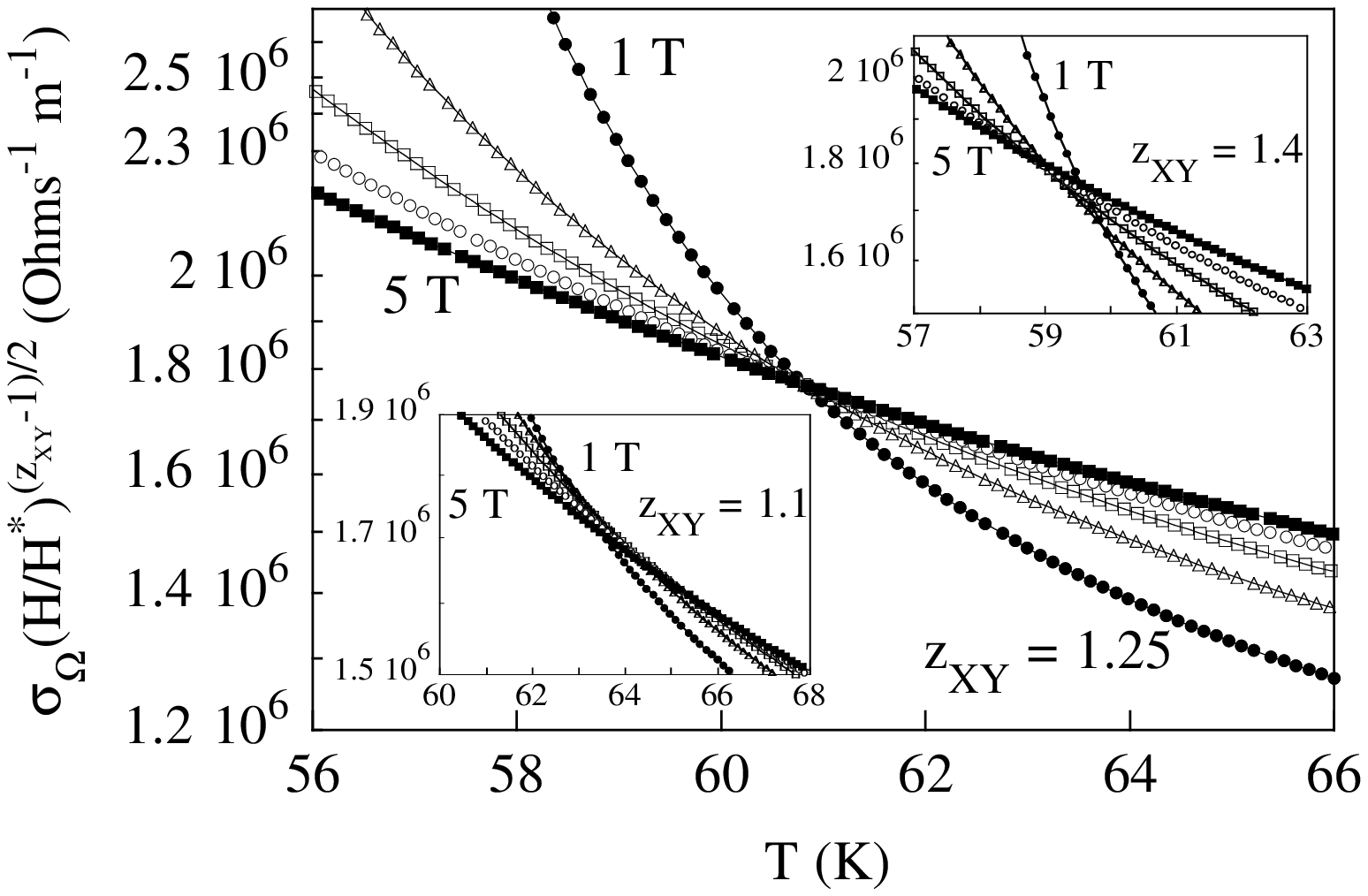}}
\end{figure}
\vspace{-.3in}
\noindent FIG.~2.
3D~$XY$ crossing point method, shown for the $\delta=0.57$ film.  
Symbols correspond to 
different fields:  1, 2, 3, 4, 5 T.  The appropriate choice of $z_{XY}$ 
causes lines to cross at a single point, identifying both $z_{XY}$ 
and 
$T_c$.  Poor crossing behavior is observed for (slightly) 
incorrect values of $z_{XY}$ (insets).  

\vspace{1.5in}
\vfill
$\mbox{}$
\vspace{-.1in}
\begin{figure}
\narrowtext
\epsfxsize=2.9truein
\vbox{\hskip 0.15truein
\epsffile{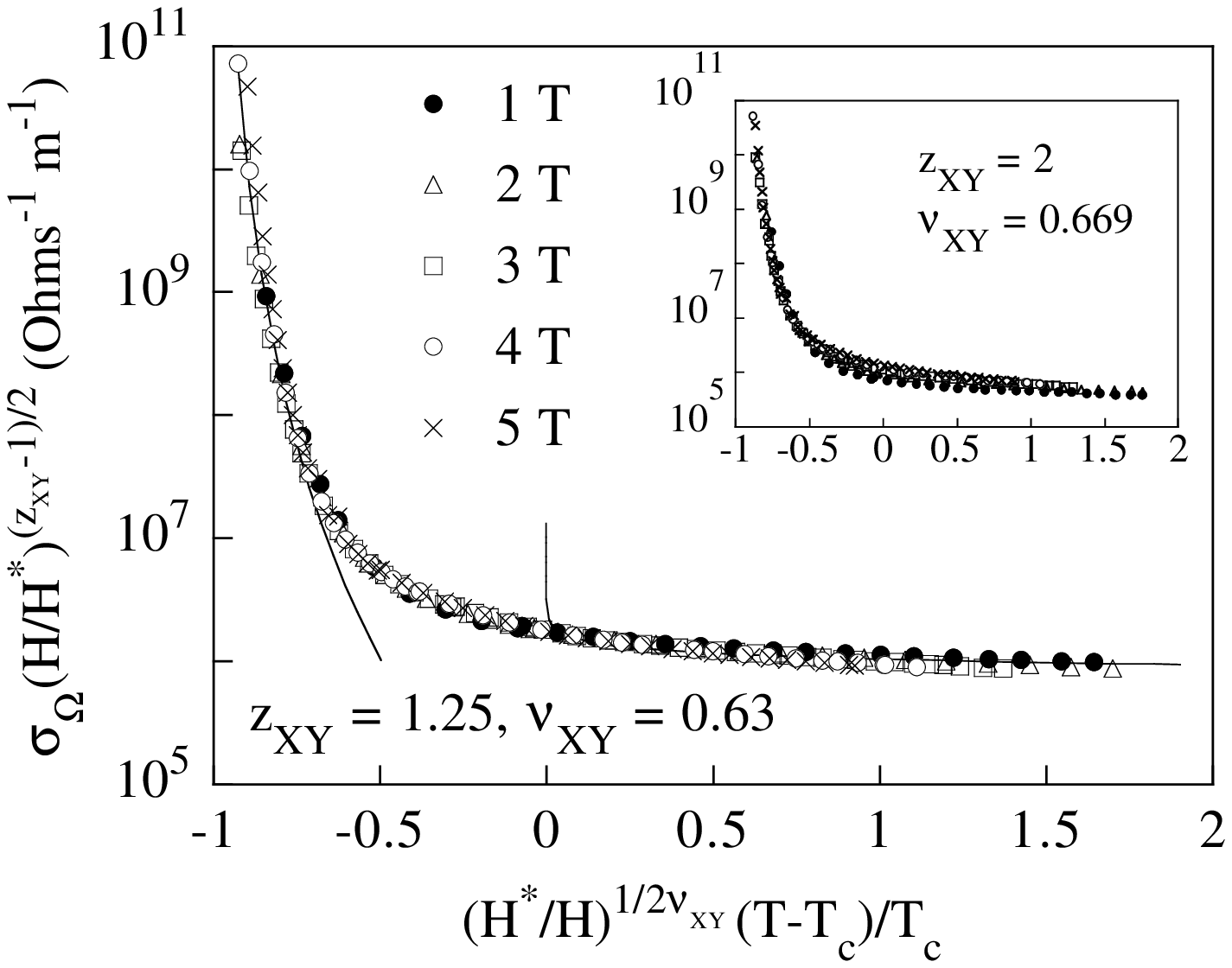}}
\end{figure}
\vspace{-.4in}
\noindent FIG.~3.
Full 3D~$XY$ scaling of the ohmic conductivity $\sigma_\Omega$ for 
the $\delta \simeq 0.57$ film.  Scaling is 
successful only for low fields ($H<H_b\simeq 5.1$ T).  
Expected asymptotic behaviors are shown as solid lines.
Insets show the poor scaling obtained
using the expected exponents $\nu_{XY}=0.669$ and $z_{XY}=2$. 

\vspace{-.15in}
\begin{figure}
\narrowtext
\epsfxsize=2.9truein
\vbox{\hskip 0.15truein
\epsffile{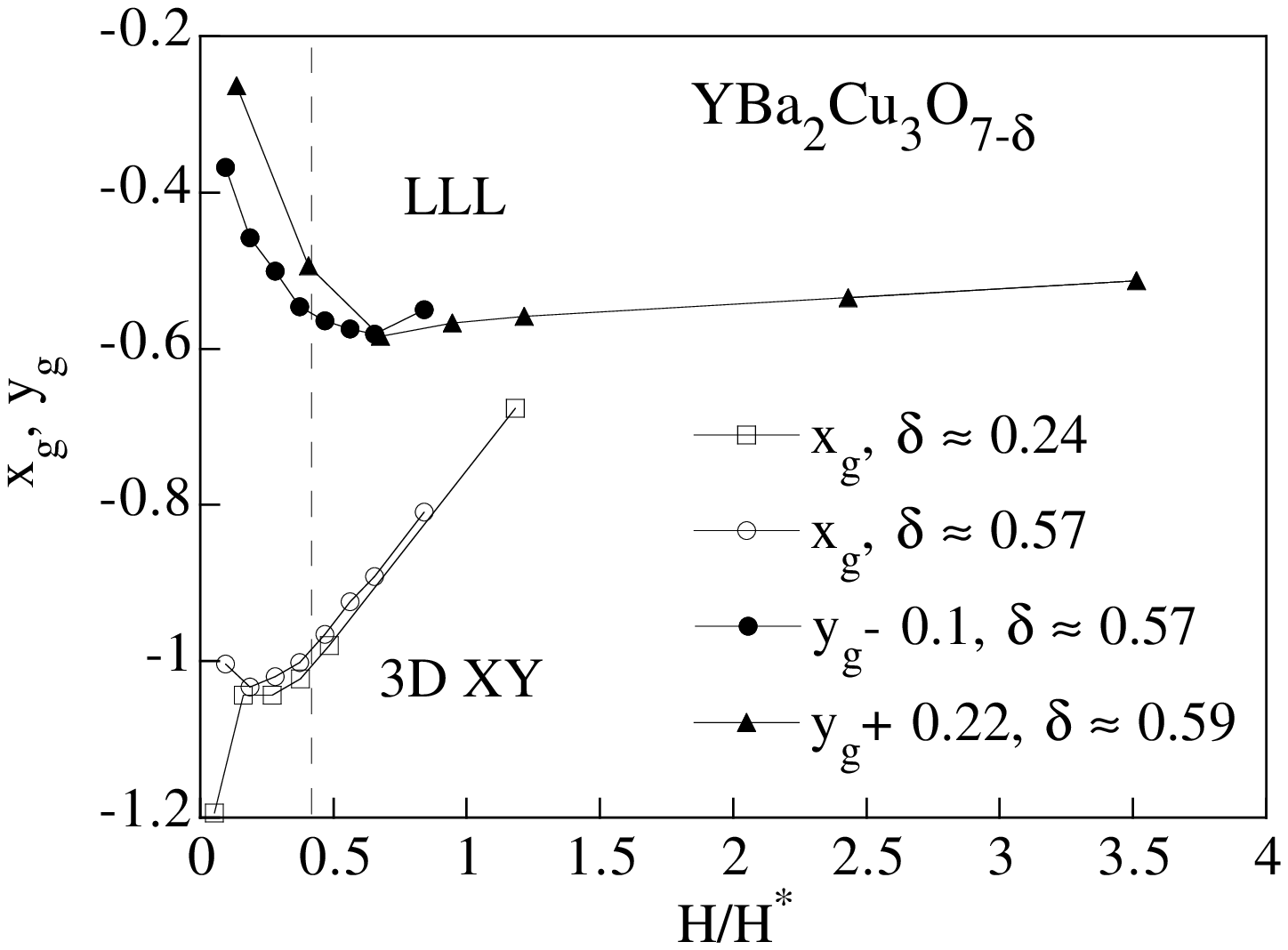}}
\end{figure}
\vspace{-.6in}
\noindent FIG.~4.
Multicritical self-consistency:  3DXY and LLL scaling parameters 
$x_g$ 
and $y_g$, 
respectively, 
evaluated at $T=T_g(H)$.  
Dashed line indicates the crossover $H_b$.  
Scaling variables should remain field-independent
in their respective scaling regions.
For $H<H_b$, 
3DXY scaling is self-consistent, while for $H>H_b$, 
LLL scaling is self-consistent.  

\vspace{-.15in}
\begin{figure}
\narrowtext
\epsfxsize=2.9truein
\vbox{\hskip 0.15truein
\epsffile{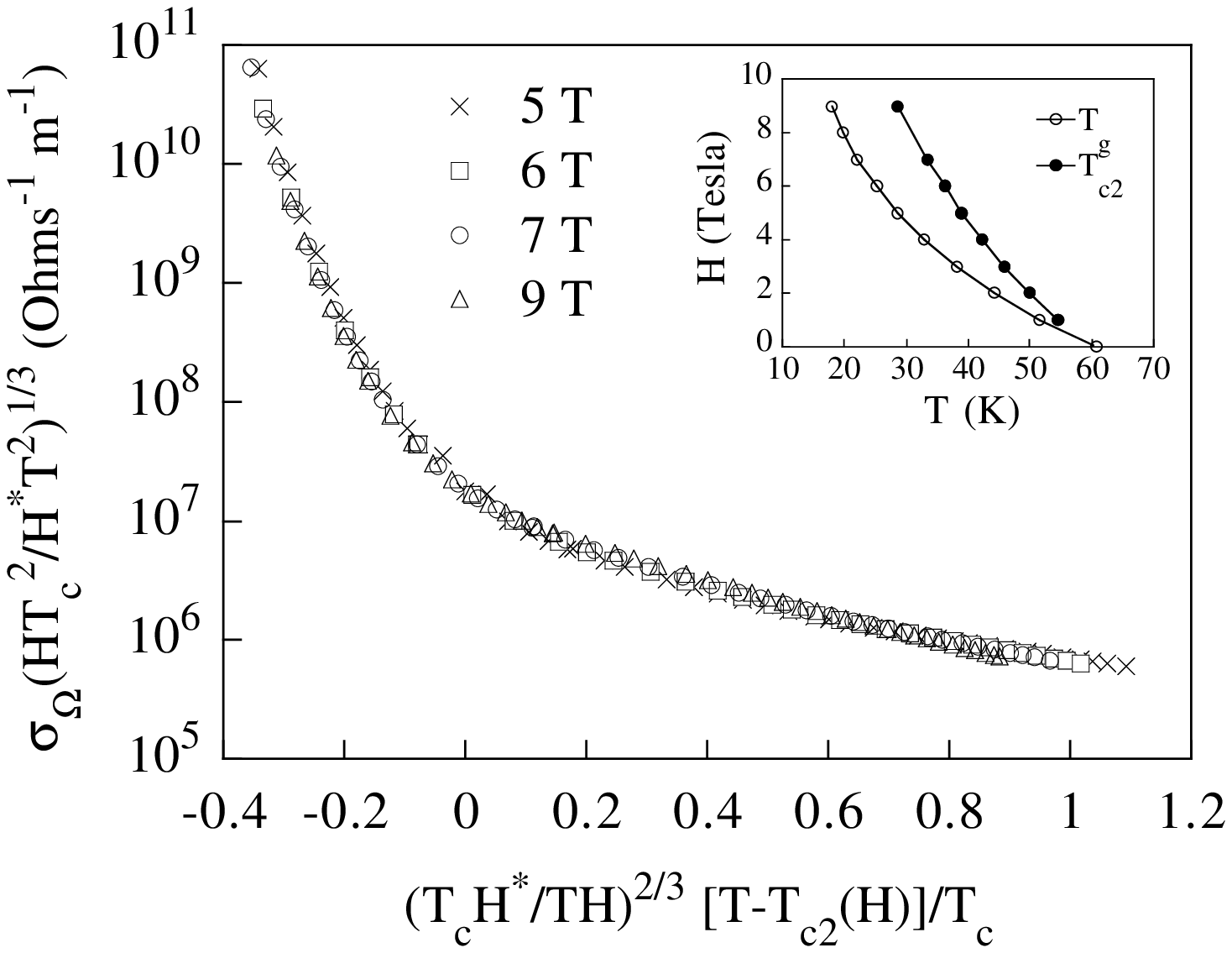}}
\end{figure}
\vspace{-.4in}
\noindent FIG.~5.
Full LLL scaling of $\sigma_\Omega$ for the 
$\delta \simeq$ 0.57 film.  Multicritically self-consistent scaling is 
successful only for high fields ($H>H_b\simeq 5.1$ T).  Inset shows 
$T_{c2}(H)$ and $T_g(H)$
for the same film.  [$T_{c2}(H)$ is speculative for $H<H_b$.] 

\end{multicols}

\begin{references}
\bibitem{lawrie}I. D. Lawrie, \prb {\bf 50}, 9456 (1994).
\bibitem{FFH}M. P. A. Fisher, \prl {\bf 62}, 1415 (1989);
	D. S. Fisher, {\it et al.,} \prb {\bf 43}, 130 (1991).
\bibitem{inderhees}S. E. Inderhees, {\it et al.,} \prl {\bf 66}, 232
	(1994);
	G. Mozurkewich, {\it et al.,} \prb {\bf 46}, 11914 (1992);
	N. Overend, {\it et al.,} \prl {\bf 72}, 3238 (1994);
	M. A. Howson, {\it et al.,} \prl {\bf 74}, 1888 (1995);
	N. Overend, {\it et al.,} \prl {\bf 75}, 1870 (1995).
\bibitem{salamon}M. B. Salamon, {\it et al.,} \prb {\bf 47}, 5520
	(1993);
	M. B. Salamon, {\it et al.,} Physica A {\bf 200}, 365 (1993).
\bibitem{hubbard}M. A. Hubbard, {\it et al.,} Physica C {\bf 259},
	309 (1996).
\bibitem{kamal}S. Kamal, {\it et al.,} \prl {\bf 73}, 1845 (1994);
	S. M. Anlage, {\it et al.,} \prb {\bf 53}, 2792 (1996).
	Note also the counterexample of 
	Z.-H. Lin, {\it et al.}, Europhys. Lett. {\bf 32}, 573 (1995).
\bibitem{yeh}N.-C. Yeh, {\it et al.,} \prb {\bf 47}, 6146 (1993);
	M. A. Howson, {\it et al.,} \prb {\bf 51}, 11984 (1995).
\bibitem{bray}A. J. Bray, \prb {\bf 9}, 4752 (1974);
	D. J. Thouless, \prl {\bf 34}, 946 (1975);
	G. J. Ruggeri and D. J. Thouless, J. Phys. F {\bf 6}, 2063 (1976);
	R. Ikeda, {\it et al.,} J. Phys. Soc. Jpn {\bf 58}, 1377 (1989);
	S. Hikami and A. Fujita, \prb {\bf 51}, 6379 (1990).
\bibitem{ullah}S. Ullah and A. T. Dorsey, \prl {\bf 65}, 2066 (1990);
	\prb {\bf 44}, 262 (1991).
\bibitem{welp}U. Welp, {\it et al.,} \prl {\bf 67}, 3180 (1991).
\bibitem{zhou}B. Zhou, {\it et al.,} \prb {\bf 47}, 11631 (1993);
	A. Junod, {\it et al.,} Physica C {\bf 211}, 304 (1993);
	S. W. Pierson, {\it et al.,} \prl {\bf 74}, 1887 (1995);
	M. Roulin, {\it et al.,} \prl {\bf 75}, 1869 (1995);
	M. Roulin, {\it et al.,} Physica C {\bf 244}, 225 (1995);
	S. W. Pierson, {\it et al.,} \prb {\bf 53}, 8638 (1996).
\bibitem{jeandupeux}O. Jeandupeux, {\it et al.,} \prb {\bf 53}, 
	12475 (1996).
\bibitem{kim}D. H. Kim, {\it et al.,} \prb {\bf 45}, 10801 (1992).
\bibitem{zlatko}Z. Te\v{s}anovi\'{c}, \prb {\bf 51}, 16204 (1995).
\bibitem{hou}L. Hou, {\it et al.,} \prb {\bf 50}, 7226 (1994).
\bibitem{zinn}J. C. LeGuillou and J. Zinn-Justin, J. Phys. (Paris) 
	{\bf 46}, L137 (1985).
\bibitem{deak}J. Deak, {\it et al.,} \prb {\bf 51}, 705 (1995).
\bibitem{schilling}A. Schilling, {\it et al.,} \prl {\bf 71}, 1899 (1993);
	C. C. Almasan and B. Maple \prb {\bf 53}, 2882 (1996);
	K. Kishio, in {\it Physical Properties of High Temperature
	Superconductors}, edited by D. M. Ginsberg (World Scientific 
	Singapore, 1996).
\bibitem{roberts}J. M. Roberts, {\it et al.,} \prb {\bf 49},
	6890 (1994).
\bibitem{friesen}M. Friesen, {\it et al.,} unpublished.
\bibitem{anlage}J. C. Booth, {\it et al.,} \prl {\bf 77}, 4438 (1996).
\bibitem{kresin}Yu. N. Ovchinnikov and V. Z. Kresin, \prb {\bf 54},
	1251 (1996), and references therein.
\bibitem{urbana}D. Ginsberg, J.-T. Kim, and N. Goldenfeld 
(unpublished).
\end{references}
\end{document}